\documentclass[twoside,11pt]{article}

\usepackage[accepted]{melba}

\usepackage{mwe} 

\usepackage{amsmath,amsfonts}

\usepackage{amsmath}
\usepackage{subfigure}
\usepackage{caption}
\usepackage{multirow}
\graphicspath{{./images/}}
\usepackage{tabularx}

\melbaid{2024:014}  
\doi{https://doi.org/10.59275/j.melba.2024-a333}
\melbaauthors{Nakarmi, Pudasaini, Thapaliya, Upretee, Shrestha, Giri, Neupane, and Khanal}  
\volume{2}
\firstpageno{956}  
\melbayear{2024}  
\datesubmitted{04/2023}  
\datepublished{08/2024}  

\melbaspecialissue{Medical Imaging with Deep Learning (MIDL) 2020}
\melbaspecialissueeditors{Marleen de Bruijne, Tal Arbel, Ismail Ben Ayed, Hervé Lombaert}

\ShortHeadings{Deep-learning Assisted Detection and Quantification of (oo)cysts of Giardia and}{Nakarmi, Pudasaini, Thapaliya, Upretee, Shrestha, Giri, Neupane, and Khanal}

\title{Deep-learning Assisted Detection and Quantification of (oo)cysts of Giardia and Cryptosporidium on Smartphone Microscopy Images}

\author{\name Suprim Nakarmi \orcid{0009-0003-7281-2729} \email {suprimnakarmi@gmail.com}\\  
	\addr Nepal Applied Mathematics and Informatics Institute for research (NAAMII), Kathmandu, Nepal \\
    \addr Center for Analytical Sciences, Kathmandu Institute of Applied Sciences (KIAS), Lalitpur, Nepal
    \AND
	\name Sanam Pudasaini \orcid{0000-0001-9782-3369} \email{sanam001@e.ntu.edu.sg} \\
\addr Center for Analytical Sciences, Kathmandu Institute of Applied Sciences (KIAS), Lalitpur, Nepal
\AND
\name Safal Thapaliya \orcid{0000-0002-4463-6700} \email{safal.thapaliya@naamii.org.np} \\
\addr Nepal Applied Mathematics and Informatics Institute for research (NAAMII), Kathmandu, Nepal
\AND
\name Pratima Upretee \email{upreteep@gmail.com}\\
\addr Nepal Applied Mathematics and Informatics Institute for research (NAAMII), Kathmandu, Nepal
\AND
\name Retina Shrestha \email{shresth.retina@gmail.com} \\
\addr Center for Analytical Sciences, Kathmandu Institute of Applied Sciences (KIAS), Lalitpur, Nepal
\AND
\name Basant Giri \orcid{0000-0003-4798-3414} \email{chembasant@gmail.com} \\
\addr Center for Analytical Sciences, Kathmandu Institute of Applied Sciences (KIAS), Lalitpur, Nepal
\AND
\name Bhanu Bhakta Neupane \orcid{0000-0003-0731-2552} \email{newbhanu@gmail.com} \\
\addr Central Department of Chemistry, Tribhuvan University, Kathmandu, Nepal
\AND
\name Bishesh Khanal \orcid{0000-0002-2775-4748} \email{bishesh.khanal@naamii.org.np}\\
\addr Nepal Applied Mathematics and Informatics Institute for research (NAAMII), Kathmandu, Nepal
}
\begin{document}

\maketitle
\setlength{\parskip}{3pt}
\begin{abstract}
The consumption of microbial-contaminated food and water is responsible for the deaths of millions of people annually.
Smartphone-based microscopy systems are portable, low-cost, and more accessible alternatives for the detection of \emph{Giardia} and \emph{Cryptosporidium} than traditional brightfield microscopes.
However, the images from smartphone microscopes are noisier and require manual cyst identification by trained technicians, usually unavailable in resource-limited settings.
Automatic detection of (oo)cysts using deep-learning-based object detection could offer a solution for this limitation.
We evaluate the performance of four state-of-the-art object detectors to detect (oo)cysts of \emph{Giardia} and \emph{Cryptosporidium} on a custom dataset that includes both smartphone and brightfield microscopic images from vegetable samples.
Faster RCNN, RetinaNet, You Only Look Once (YOLOv8s), and Deformable Detection Transformer (Deformable DETR) deep-learning models were employed to explore their efficacy and limitations.
Our results show that while the deep-learning models perform better with the brightfield microscopy image dataset than the smartphone microscopy image dataset, the smartphone microscopy predictions are still comparable to the prediction performance of non-experts. Also, we publicly release brightfield and smartphone microscopy datasets with the benchmark results for the detection of \emph{Giardia} and \emph{Cryptosporidium}, independently captured on reference (or standard lab setting) and vegetable samples. Our code and dataset are available at \url{https://zenodo.org/records/12587799} and \url{https://zenodo.org/records/7813183}, respectively.
\end{abstract}


\section{Introduction}
	\label{sec:introduction}
Pathogen contamination of food and water is a serious global challenge.
About $1.7$ billion people do not have access to feces-free drinking water in the world\footnote{\url{https://www.who.int/news-room/fact-sheets/detail/drinking-water}}.
Such contaminated food and water can often lead to diarrhea.
According to the World Health Organization (WHO), diarrheal diseases account for the loss of around half a million human lives annually.
Similarly, nearly $600$ million people fall ill due to the consumption of contaminated food, resulting in more than four hundred thousand deaths every year\footnotemark{}.
\emph{Giardia} and \emph{Cryptosporidium} are two of the major causes of protozoan-induced diarrheal diseases and are the most frequently identified protozoan parasites causing outbreaks \citep{baldursson2011waterborne}\footnotemark[\value{footnote}]. \footnotetext{\url{https://www.who.int/news-room/fact-sheets/detail/food-safety}}
In low- and middle-income countries, children under three years of age experience three episodes of diarrhea on average every year\footnote{\url{https://www.who.int/news-room/fact-sheets/detail/diarrhoeal-disease}}.
\emph{Giardia} and \emph{Cryptosporidium} cause intestinal illness called giardiasis and cryptosporidiosis, respectively.
Infections due to these parasites are more common in low- and middle-income countries because of unhygienic lifestyles and poor sanitation \citep{fricker2002protozoan, gupta2020prevalence, tandukar2013intestinal, sherchand2004cryptosporidium}.
\emph{Giardia} is more prominent in size with an elliptical shape having a major axis 8-12 \textmu m and minor axis 7-15 \textmu m.
In contrast, \emph{Cryptosporidium} is spherically shaped, having a diameter of 3-6 \textmu m \citep{dixon2011protozoan}.
Accurate detection of these microorganisms could enable early diagnosis, saving millions of lives.
Moreover, regularly screening these pathogens in real food and water samples could help prevent infections and disease outbreaks.

Polymerase Chain Reaction (PCR), immunological assays, cell culture methods, fluorescence in situ hybridization, and microscopic analysis \citep{van2015comparison, adeyemo2018methods} are the primary methods for the detection of (oo)cysts of \emph{Giardia} and \emph{Cryptosporidium}.
Even though these methods are reliable and accurate, they are laborious, time-consuming, costly, and require significant expertise, resulting in a lack of tests in many resource-limited regions.
For instance, cell culture requires more than $24$ hours, and immunological assays and PCR reactions require costly reagents and equipment \citep{guerrant1985evaluation, guerrant2001practice}. 
In addition, some viable bacterial pathogens are difficult to grow or are even non-culturable \citep{oliver2005viable}.
Fluorescence tagging of cells requires expertise, and the cost of reagents is high. 
Microscopic methods are the most widely used diagnostic methods for detecting parasites, especially in low-resource countries.
The microscopic method examines a glass slide containing a sample under a microscope.
However, the traditional microscopes are still costly, less portable, and require expertise to handle and accurately identify microorganisms on the slides \citep{chavan2014malaria}. 
Recently, smartphone-based microscopic methods have been developed to potentially replace or supplement more expensive and less portable microscopic methods, including the brightfield and fluorescence microscopy \citep{koydemir2015rapid,kobori2016novel,kim2015smartphone,saeed2018smart,shrestha2020smartphone,feng2016high}. These microscopes allow magnification of microorganisms enabling the user to observe them on the phone screen immediately and capture the images and videos using the smartphone.
However, smartphone microscopy also requires a well-trained person to identify target organisms accurately, analyze the result, and report it.
In the least developed countries, the lack of skilled technicians limits the use of such a new microscopic system for rapid field testing and clinical applications.
Therefore, robust and automated detection of microorganisms using smartphone microscopic images could enable more widespread use of smartphone microscopes in large-scale screening of parasites. 
\setlength{\parskip}{0pt}

In recent years, several deep learning-based algorithms have been developed for various biological and clinical applications such as automatic detection, segmentation, and classification of human cells and fungi species \citep{xue2017cell, zielinski2020deep}, bacteria \citep{wang2020early}, malaria detection \citep{vijayalakshmi2020deep}, image segmentation of two-dimensional materials in microscopic images \citep{masubuchi2020deep}.
In addition, detecting a pollen grain from microscopy images using a deep neural network has been reported \citep{gallardo2019precise}.
\cite{de2020automated} automated the screening of sickle cells using deep learning on a smartphone-based microscope. 
Similarly, \cite{xu2020deep} proposed ParasNet - a deep-learning-based network - to detect \emph{Giardia} and \emph{Cryptosporidium} in brightfield microscopic images.
\cite{luo2021deep} created MCellNet to classify \emph{Giardia}, \emph{Cryptosporidium}, microbeads, and natural pollutants from the images captured from imaging flow cytometry.
Machine learning techniques have also been developed to classify \emph{Giardia} from other parasites using features such as area, equivalent diameter, and intensity in fluorescent smartphone-based microscopic images \citep{koydemir2015rapid}.
Several other studies have proposed deep learning-based algorithms to automate microorganism detection in smartphone microscopes \citep{de2020automated,fuhad2020deep,yang2019deep}.
However, it is not clear how well state-of-the-art deep learning models perform in automated detection of the (oo)cysts of \emph{Giardia} and \emph{Cryptosporidium} from smartphone microscopic images in comparison to the traditional microscope and non-experts. We also introduce labeled smartphone and brightfield microscopy datasets for detection of the (oo) cysts of \emph{Giardia} and \emph{Cryptosporidium}, making it publicly available for the scientific community.   

Here, we explore the potential of deep learning algorithms to detect cysts of two different kinds of enteric parasites without human experts' involvement in images taken using a custom-built smartphone microscopy system. We created a custom dataset by capturing images of sample slides from a smartphone microscope and a traditional brightfield microscope. The sample slides were obtained for (oo)cysts on both reference and vegetable sample extracts. 
We trained four popular object detection models - Faster RCNN \citep{ren2016faster}, RetinaNet \citep{lin2017focal}, YOLOv8s \citep{yolov8s2023}, and Deformable DETR \citep{zhu2020deformable} - using our dataset and evaluated the performance of the models.
Finally, we report the performance of these models compared with expert humans and non-expert humans and inter-operator variability.
The performance of the models with smartphone microscopic images was also compared with brightfield microscopic images.
The dataset and the source code are made publicly available.

\section{Methods}
\label{sec:method}

\subsection{Dataset}
\label{sec:dataset}

\subsubsection{Training-Validation Set}
\label{sussec:training-validation}

We used two microscopes to capture images: i) a traditional brightfield microscope and ii) a sapphire ball lens-based smartphone microscope developed by \cite{shrestha2020smartphone}. 
The brightfield microscope images were captured with a rectangular Field Of View (FOV) of ~190 {\textmu}m X $350$ {\textmu}m and a magnification of 400X. In contrast, the smartphone microscope images were captured with a circular FOV of diameter $200$ {\textmu}m and magnification of 200X - using Samsung Galaxy J7 Prime.

We captured images by making microscope slides from two different types of samples: i) standard or reference (oo)cyst suspension samples (Waterborne Inc, PC101 G/C positive control), and ii) actual vegetable samples obtained from local markets in Nepal.
\autoref{fig:example_images} shows a few examples of reference and actual vegetable samples along with their bounding box annotations.
Likewise, \autoref{tab:dataset_info} summarizes the number and types of images captured in the dataset.

\textbf{Reference Samples:} We prepared $25$ slides each from $5$ {\textmu}L standard (oo)cyst suspension (or standard samples) mixed with Lugol's iodine in equal proportion. Slides refer to microscope slides on which samples were mounted, generally used for microscopic examinations.
From these slides, an expert with more than two years of experience in imaging and annotating \emph{Giardia} and \emph{Cryptosporidium} from microscopy images captured many images from both microscopes.
To maintain consistency, we selected $830$ best images, each based on the clarity of parasites. 
Images with one of the following criteria were removed: (i) images consisting of only broken parasites - either completely damaged or partially captured, and (ii) Noisy or blurry images where no single object was visible. 
Since there was no prior knowledge of parasite counts in individual slides, identification and counting of (oo)cysts by the expert was considered as a benchmark.

\textbf{Vegetable Samples:} We prepared $200$ slides from $151$ vegetable samples collected from local markets in Nepal.
The expert captured and selected images using the same protocol as the one used for
the reference samples. 
From these slides, 1005 images for each type of microscope were included for further steps. 
The images were used for training the object detection models.

\subsubsection{Independent Test Set}
\label{sec:independent-testset}
To assess the generalization of deep learning models on smartphone microscope images, an independent test set was prepared as follows:
One hundred ninety-three images were captured on a particular day using the same smartphone microscope. The expert captured images at random locations of the microscope slides, regardless of the presence of parasites. 
The expert annotated these images with bounding boxes and ellipses for all \emph{Giardia} and \emph{Cryptosporidium}  using VGG annotator \citep{dutta2019vgg}. 

\subsubsection{Non-Expert Annotated Set}
\label{sec:non-expert}
The same expert trained three non-expert humans on smartphone vegetable sample images for three hours in a single-day training session. 
As the shape and size of (oo)cysts of \emph{Giardia} and \emph{Cryptosporidium} played a crucial role in their identification, the non-experts were instructed to use these features to locate the parasites on images.
After the training session, the three non-experts annotated the cysts in $193$ independent test set images. 
Note that non-experts used MS-paint for annotations by encircling the cysts of the two parasites with ellipses of two different colors (yellow: \emph{Giardia}, and blue: \emph{Cryptosporidium}), and the time non-experts took to annotate the cysts in each of the images were recorded using a stopwatch.

\begin{figure*}[htb]
\centering
 \includegraphics[width=1\textwidth, scale=0.8]{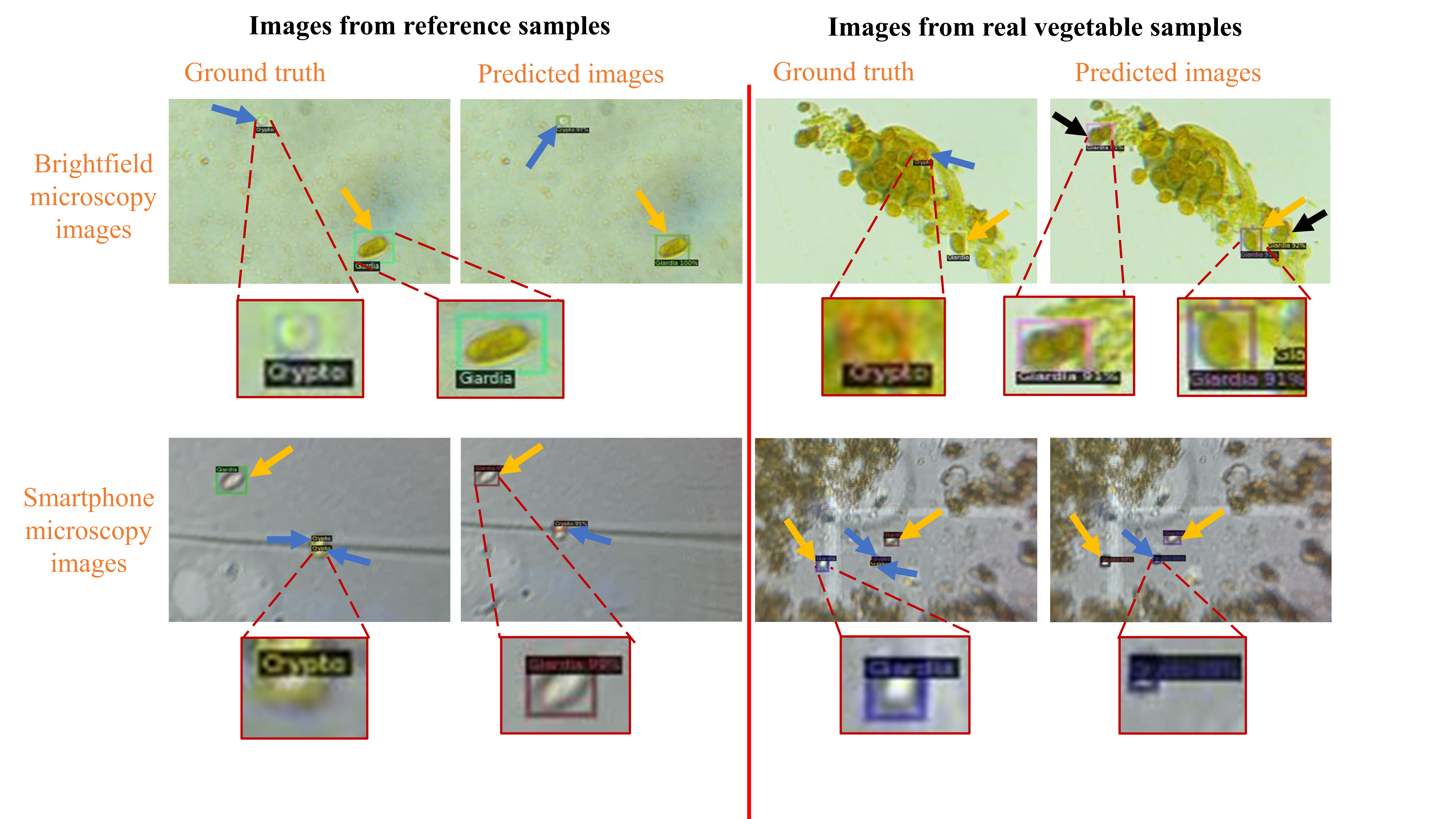}
\caption{ Representative images of reference and real vegetable samples captured using brightfield and smartphone microscopes. Example ground truth annotations of the parasites by the expert and predictions from the object detector models. Yellow arrow points \emph{Giardia}, blue arrow points \emph{Cryptosporidium}, and black arrow points debris.  For this illustration, Faster RCNN was used to detect parasites.} 
\label{fig:example_images}
\end{figure*}

\begin{table*}[htb!]
  \centering
  \caption{| Description of two types of the dataset (reference sample and vegetable sample) used in this study with the total number of parasites annotated by the expert using VGG software \citep{dutta2019vgg}.}
    \resizebox{\columnwidth}{!}{%
    \begin{tabular}{lclcc} 
    
    \hline
    
    \bf{Dataset} & \multicolumn{1}{l}{\bf{No. of images}} & \bf{Microscopes} & \bf{Giardia's annotation} & \bf{Cryptosporidium's annotation} \\
    
    \hline
    
    Reference sample & $830$  & Smartphone & $839$  & $534$ \\
          &       & Brightfield & $907$  & $502$ \\
    
    \hline
    
    Vegetable sample & $1005$ & Smartphone & $439$  & $796$ \\
          &       & Brightfield & $344$  & $740$ \\
    
    \hline
    
    Test (vegetable sample)  & $193$  & Smartphone & $165$  & $137$ \\
    
    \hline
    \end{tabular}%
    }
  \label{tab:dataset_info}
\end{table*}

\subsection{Deep Learning-based Object Detection Models}
\label{sec:models}
Four state-of-the-art object detection models were selected for this study: Faster RCNN \citep{ren2016faster}, RetinaNet \citep{lin2017focal}, YOLOv8s \citep{yolov8s2023}, and Deformable DETR \citep{zhu2020deformable}.
Faster RCNN is one of the most popular networks from the Region-Based Convolutional Neural Networks (RCNN) family, which is an improved version based on two previous methods, RCNN \citep{girshick2014rich} and Fast RCNN \citep{girshick2015fast}.
RetinaNet is a popular single-stage detector that uses a \textit{Focal Loss} to address the foreground-background class imbalance problem that gets more severe when detecting smaller objects in the images. 
YOLOv8s is a popular single-stage detector with relatively low computational costs, enabling it to be run on smartphones. 
To assess the quality of the real-time detection model that can be run on smartphones, we chose YOLOv8s - the smallest model with $11.2$ million parameters - from the various available models for YOLOv8, as it is lightweight and has the lowest prediction time \citep{yolov8s2023}. Deformable DETR is a state-of-the-art transformer-based object detector that incorporates deformable attention mechanisms, replacing the fixed grids used in traditional self-attention with deformable grids. This allows the model to capture spatial relationships more effectively by dynamically adjusting its attention \citep{zhu2020deformable}, which is the improved version of DETR.

\subsection{Experimental Setup}
\label{sec:experimental_setup}
\autoref{fig:block_diagram} illustrates the overall pipeline of the training and evaluation of the four object detectors using expert and non-expert annotated images captured from brightfield and smartphone images.
During training, the images were first pre-processed (see \autoref{sec:imp_details}) and then fed as an input to one of the four object detection models for classifying the object type (\emph{Giardia} or \emph{Cryptosporidium}) and localizing the objects with a bounding box.

\begin{figure*}[htb]
    \centering
    \includegraphics[width=\textwidth]{images/schema.png}
    \caption{Schema showing the overview of the method implemented in this study. Note that the detector networks were trained separately using images captured from brightfield and smartphone microscopes on reference and vegetable samples. The testing phase consists of predicting (oo)cysts of the parasites only on images captured on vegetable samples using the smartphone microscope. } 
    \label{fig:block_diagram}
\end{figure*}

\subsubsection{Evaluation Approach}
Since the target application is to be able to assess the contamination in vegetable samples by identifying and counting the number of (oo)cysts of the two parasites in the microscopic images, we evaluate the four models using classification performance metrics: precision, recall, and F1-score.
The model-predicted cysts can belong to one of the three categories: True Positive (TP) when the model's object prediction (of either \emph{Giardia} or \emph{Cryptosporidium} cysts) correctly matches with the Ground Truth (GT) annotation, False Positive (FP) when the model's object prediction is different than the GT, and finally False Negative (FN) when the model does not detect cysts annotated by the experts in the images.
Precision, Recall, and F1-scores are calculated based on these three values to evaluate the models' performances.
\begin{equation}
\label{eqn:precision}
    Precision \ (P) = \frac{TP}{TP + FP}
\end{equation}

\begin{equation}
\label{eqn:recall}
    Recall \ (R) = \frac{TP}{TP +  FN}
\end{equation}

\begin{equation}
\label{eqn:f1score}
    F1 \ \- score = \frac{2 \times P \times R}{P + R}
\end{equation}

In addition to assessing the object detection model's performance against the expert GT annotations, we also evaluate how the model compares with non-expert humans.
This helps assess the utility of deploying the automated models in places where the experts are not available.

\subsubsection{Comparing the Four Models and Non-expert Humans} 
We evaluated the object detection models - Faster RCNN, RetinaNet, YOLOv8s, and Deformable DETR - in the following settings:
\begin{itemize}
    \item 5-fold cross-validation of the brightfield and smartphone microscope images using the training-validation dataset with expert annotations: $830 \times 2$ images for the two microscopes with reference and $1005 \times 2$ images with vegetable samples.
    \item Comparison of the detection models against non-expert humans in identifying the (oo)cysts of the two parasites in separate independent smartphone microscope test images ($n=193$).
    \item Comparison of time taken by non-experts vs. detection model to identify the (oo)cysts in the smartphone microscope test images.
\end{itemize}

To test whether the differences in the performance of the non-expert humans and the detection models in the independent test set are statistically significant, we use paired Wilcoxon signed-rank test \citep{woolson2007wilcoxon}. The details are provided in \autoref{sec:stat}.

\section{Results}
\label{sec:results}

\subsection{5-fold Cross-validation in Training-validation Set} 
\autoref{tab:crossval-brightfield} presents the performance of the four object detection models for the brightfield microscope images of reference and vegetable samples.
The results are reported for confidence scores ($c$) and iou-thresholds ($i$) optimized for each model separately.
The details about the choice of these thresholds are provided in \autoref{sec:thresholds_choice}.
We see that the models perform better for reference samples than vegetable samples.
It is expected because the reference sample does not have debris, and hence, there are very few objects confounding with the cysts in the clean brightfield images (example images in \autoref{fig:example_images}).
Similarly, we observe that all the object detectors detect \emph{Giardia} cysts better than \emph{Cryptosporidium} ones, except for Faster RCNN, which has negligibly better performance in detecting \emph{Cryptosporidium}.

\begin{table*}[htb!]
\centering
\caption{Brightfield microscope training-validation set: 5-fold cross-validation results on reference and vegetable samples for Faster RCNN, RetinaNet, YOLOv8s and Deformable DETR.}
\label{tab:crossval-brightfield}

\resizebox{\columnwidth}{!}{%
\begin{tabular}{lllccc}
\hline 
\bf{Dataset} & \bf{Models}  &\bf{Thresholds}  &\bf{Precision}  &\bf{Recall} &\bf {F1-score}\\
\hline 
 \multicolumn{6}{c}{\emph{Giardia}}\\

\hline  

Reference Sample & Faster RCNN & $c=0.7,i=0.4$ & $\textbf{0.957} \pm 0.014$ & $0.968 \pm 0.018$ & $0.962 \pm 0.007$ \\
                 & RetinaNet & $c=0.5,i=0.5$ & $0.917 \pm 0.017$ & $0.946 \pm 0.013$ & $0.931 \pm 0.011$\\ 
                 & YOLOv8s & $c=0.4,i=0.3$ & $0.954 \pm 0.016$ & $\textbf{0.975} \pm 0.010$ & $\textbf{0.965} \pm 0.010$ \\
                  & Deformable DETR &  $c=0.4,i=0.2$ & $0.864 \pm 0.037$ & $0.870 \pm 0.014$ & $0.867 \pm 0.021$\\
                 
\hline

Vegetable Sample & Faster RCNN  & $c=0.8,i=0.5 $ & $0.783 \pm 0.024$ & $0.901 \pm 0.026$   & $ \textbf{0.837} \pm 0.005 $ \\            
                & RetinaNet & $c=0.4,i=0.4$ & $0.753 \pm 0.029$ & $\textbf{0.927} \pm 0.035$ & $0.830 \pm 0.024$\\ 
                & YOLOv8s & $c=0.4,i=0.3$ & $\textbf{0.855} \pm 0.037$ & $0.811 \pm 0.018$ & $0.831 \pm 0.018$ \\
                & Deformable DETR &  $c=0.4,i=0.2$ & $0.636 \pm 0.042$ & $ 0.782 \pm 0.063$ & $0.700 \pm 0.042$ \\
\hline
 \multicolumn{6}{c}{\emph{Cryptosporidium}}\\
\hline
Reference Sample & Faster RCNN & $c=0.7,i=0.4$ & $0.880 \pm 0.024$ & $\textbf{0.915} \pm 0.030$ & $\textbf{0.897} \pm 0.018$ \\       
                & RetinaNet & $c=0.5,i=0.5$ & $\textbf{0.917} \pm 0.025$ & $0.870 \pm 0.031$ & $0.893 \pm 0.022$ \\         
                & YOLOv8s & $c=0.4,i=0.3$ & $0.890 \pm 0.034$ & $0.887 \pm 0.031$ & $0.888 \pm 0.028$ \\
                & Deformable DETR &  $c=0.4,i=0.2$ & $0.768 \pm 0.035$ & $ 0.808 \pm 0.012$ & $0.787 \pm 0.015$ \\
\hline
Vegetable Sample & Faster RCNN & $c=0.8,i=0.5$ & $0.845 \pm 0.029$ & $0.835 \pm 0.045$   &  $\textbf{0.839} \pm 0.024$  \\  
                & RetinaNet & $c=0.4,i=0.4$ & $0.801 \pm 0.025$ & $\textbf{0.851} \pm 0.022$ & $0.826 \pm 0.023$ \\      
                & YOLOv8s & $c=0.4,i=0.3$ & $\textbf{0.879} \pm 0.052$ & $0.716 \pm 0.065$ & $0.788 \pm 0.051$ \\
                & Deformable DETR &  $c=0.4,i=0.2$ & $0.642 \pm 0.029$ & $0.720 \pm 0.054$ & $0.678 \pm 0.033$\\
 \hline            

\end{tabular}%
}
\end{table*}

Similarly, \autoref{tab:crossval-smartphone} shows 5-fold cross-validation results for smartphone microscope training-validation set images.
As expected, the object detectors' performance is lower than brightfield microscope images because smartphone microscope images have more textured noise and lower magnification than traditional brightfield microscopes. However, Deformable DETR performs slightly better when predicting \emph{Cryptosporidium} on vegetable samples.
We see that most of the trends observed in brightfield microscope images can also be seen for smartphone microscope:
\emph{Giardia} cysts are better detected than \emph{Cryptosporidium} cysts, YOLOv8s and Faster RCNN provide better results than RetinaNet, except for vegetable samples, and the performance of the detectors is better for reference samples compared to the vegetable sample in general.
However, one notable exception is that the F1-score for the vegetable sample is better than the reference sample for \emph{Cryptosporidium} when using YOLOv8s and Deformable DETR.
In the case of YOLOv8s, while the recall reduced slightly in the vegetable sample, the precision increased substantially, increasing the overall F1-score. 
Similarly, a remarkable increase in precision and recall is observed in Deformable DETR. 
RetinaNet provides better results in detecting \emph{Giardia} and \emph{Cryptosporidium} in vegetable samples, but Faster RCNN is better for reference samples.

\begin{table*}[htb!]
\centering
\caption{Smartphone microscope training-validation set: 5-fold cross-validation results on reference and vegetable samples for Faster RCNN, RetinaNet, YOLOv8s, and Deformable DETR.}
\label{tab:crossval-smartphone}
\resizebox{\columnwidth}{!}{%
\begin{tabular}{llcccc}

\hline 
\bf{Dataset} & \bf{Models}  &\bf{Thresholds}  &\bf{Precision}  &\bf{Recall} &\bf {F1-score}\\
\hline 

 \multicolumn{6}{c}{\emph{Giardia}}
\\ 
\hline

Reference Sample & Faster RCNN & $c=0.5,i=0.3$ & $\textbf{0.872} \pm 0.038$ & $\textbf{0.919} \pm 0.019$ & $\textbf{0.895} \pm 0.024$\\
                & RetinaNet & $c=0.3,i=0.3$ & $0.803 \pm 0.024$ & $0.912 \pm 0.022$ & $0.854 \pm 0.018$ \\
                & YOLOv8s & $c=0.3,i=0.4$ & $0.835 \pm 0.032$ & $0.891 \pm 0.032$ & $0.862 \pm 0.030$ \\
                  & Deformable DETR &  $c=0.4,i=0.2$ & $0.768 \pm 0.038$ & $ 0.745 \pm 0.022$ & $0.756 \pm 0.027$ \\
                 \hline

Vegetable Sample & Faster RCNN & $c=0.5,i=0.4$ & $0.675 \pm 0.061$ & $\textbf{0.789} \pm 0.047$ & $0.726 \pm 0.045$\\
                & RetinaNet & $c=0.4,i=0.3$ & $0.729 \pm 0.040$ & $0.751 \pm 0.061$ & $0.739 \pm 0.042$\\
                & YOLOv8s & $c=0.2,i=0.3$ & $\textbf{0.761} \pm 0.053$ & $0.729 \pm 0.093$ & $\textbf{0.740} \pm 0.038$ \\
                & Deformable DETR &  $c=0.4,i=0.2$ & $0.700 \pm 0.025$ & $0.690 \pm 0.039$ & $0.694 \pm 0.023$ \\
                 \hline             
  \multicolumn{6}{c}{\emph{Cryptosporidium}}\\
\hline 

Reference Sample & Faster RCNN & $c=0.5,i=0.3$ & $\textbf{0.761} \pm 0.030$ & $0.673 \pm 0.051$ & $\textbf{0.714} \pm 0.037$ \\
                & RetinaNet & $c=0.3,i=0.3$ & $0.698 \pm 0.027$ & $\textbf{0.688} \pm 0.045$ & $0.692 \pm 0.025$\\
                & YOLOv8s & $c=0.3,i=0.4$ & $0.580 \pm 0.059$ & $0.678 \pm 0.117$ & $0.620 \pm 0.066$ \\
                & Deformable DETR &  $c=0.4,i=0.2$ & $0.580 \pm 0.038$ & $0.598 \pm 0.058$ & $0.588 \pm 0.042$ \\
\hline

Vegetable Sample & Faster RCNN & $c=0.5,i=0.4$ & $0.638 \pm 0.031$ & $0.650 \pm 0.045$ & $0.644 \pm  0.030$ \\
                & RetinaNet & $c=0.4,i=0.3$ & $\textbf{0.700} \pm 0.046$ & $0.675 \pm 0.024$ & $\textbf{0.686} \pm 0.017$ \\
                & YOLOv8s & $c=0.2,i=0.3$ & $0.658 \pm 0.021$ & $0.676 \pm 0.103$ & $0.663 \pm 0.046$ \\
                & Deformable DETR &  $c=0.4,i=0.2$ & $0.675 \pm 0.034$ & $ \textbf{0.695} \pm 0.038$ & $0.684 \pm 0.029$ \\
\hline  
\end{tabular}%
}
\end{table*}

\subsection{Independent Test Set with Smartphone Microscope Images}
\autoref{tab:test-set} presents the performance of the four models and non-expert humans on the independent test set images.
This set consists of smartphone microscopic images of vegetable samples where the expert's annotation is considered ground truth.
The performance of all the models on the test set images is lower than that of cross-validation, which is typical for machine learning models, known as the generalization problem.
YOLOv8s, which performed well in cross-validation, has the worst overall score in the test set, suggesting that the model was less robust to new data.
In contrast, the RetinaNet model is more robust to the new data.
Moreover, RetinaNet seems to perform better for \emph{Cryptosporidium}, which could be due to the \textit{Focal Loss} targeted for smaller objects, as \emph{Cryptosporidium} cysts have a smaller size than \emph{Giardia} cysts.
Nevertheless, the test set results reinforce the observation that detecting \emph{Cryptosporidium} is more difficult compared to \emph{Giardia} cysts.

\begin{table*}[htb!]
    \small
    \centering
    \caption{Smartphone microscope independent test set: Performance of the four object detectors and non-expert humans. The standard deviation describes how diverse the scores are from the mean when predicting the test set on each cross-validation fold.  
    }
    \resizebox{\columnwidth}{!}{%
    \begin{tabular}{lccc|ccc}
   \multirow{2}{*}{\textbf{Detector}}& \multicolumn{3}{c}{\textbf{\emph{Giardia}}} & \multicolumn{3}{c}{\textbf{\emph{Cryptosporidium}}}\\
    \cline{2-7}
    & \textbf{Precision} & \textbf{Recall} & \textbf{F1-score} & \textbf{Precision} & \textbf{Recall} & \textbf{F1-score}\\
    \hline
    Faster RCNN & $\textbf{0.679} \pm 0.024$ & $\textbf{0.358} \pm 0.015$  & $\textbf{0.468} \pm 0.015$  & $0.423 \pm 0.030$ & $0.270 \pm 0.033$& $0.328 \pm 0.026$  \\
    RetinaNet & $0.657 \pm 0.034$ & $0.333 \pm 0.025$ & $0.441 \pm 0.025$ & $\textbf{0.473} \pm 0.013$ & $0.286 \pm 0.049$ & $\textbf{0.355} \pm 0.039$ \\ 
    YOLOv8s & $0.656 \pm 0.020$ & $0.240 \pm 0.037$ & $0.350 \pm 0.040$ & $0.337 \pm 0.019$ & $0.260 \pm 0.037$ & $0.293 \pm 0.031$ \\ 
     Deformable DETR & $0.578 \pm 0.018$ & $0.272 \pm 0.036$ & $0.368 \pm 0.034$ & $0.350 \pm 0.019$ & $\textbf{0.340} \pm 0.012$ & $0.344 \pm 0.009$ \\
    \hline
    Non-expert1 & $0.634$ & $0.430$ & $0.512$ & $0.233$ & $0.124$ & $0.162$ \\
    Non-expert2 & $0.399$ & $0.539$ & $0.459$ & $0.191$ & $0.606$ & $0.290$ \\
    Non-expert3 & $0.297$ & $0.497$ & $0.372$ & $0.168$ & $0.255$ & $0.203$ \\
    \hline
    Expert (Benchmark) & $1.000$ & $1.000$ & $1.000$ & $1.000$ & $1.000$ & $1.000$  \\
    \hline
    \end{tabular}%
    }
   
    \label{tab:test-set}
\end{table*}

Wilcoxon signed-rank test showed that all the models and non-expert humans mostly had significantly different results, except for Non-expert $1$, who had non-significant statistical results with Faster RCNN and RetinaNet, for both \textit{Giardia} and \textit{Cryptosporidium}.
(Details provided in \autoref{sec:stat}.)

Since the ability to detect the cysts in real-time using only the smartphone can be valuable for rapid field testing scenarios, we computed the time required to detect the cysts for the different models and the human experts and non-experts.
YOLOv8s was the fastest, predicting the objects in an average of $0.032$ seconds per image, whereas Faster RCNN, RetinaNet, and Deformable DETR needed $1.4$ seconds, $1.3$ seconds, $0.25$ seconds, respectively. Note that all the models were trained and predicted on Google Colab having Central Processing Unit (CPU) and Graphics Processing Unit (GPU) specifications of Intel (R) Xeon (R) 2vCPU @ 2.2 GHz and Tesla T4 (16GB, 2560 CUDA cores), respectively. 
Among the human annotators, the expert identified cysts in an average of $8.4$ seconds per image, but the non-experts took as long as $24.818$ seconds.

\section{Discussion and Conclusion}
\label{sec:discussion} 
In this study, we explored the possibility of using automatic parasite detection in brightfield and smartphone microscopes for the (oo)cysts of \emph{Giardia} and \emph{Cryptosporidium} in scenarios where experts are not available. 
Two different datasets were prepared by separately capturing the images of reference and actual vegetable samples using the smartphone and the brightfield microscopes. 
Four object detection models were explored, and their performance was compared against human non-experts while taking an expert annotation as ground truth. 
Precision, recall, and F1-scores were used as they are useful evaluation metrics when a target application requires counting objects \citep{xue2017cell,brhane2019concorde}.

\begin{figure*}[htb]
    \centering
    \includegraphics[width=1\textwidth, scale=0.2]{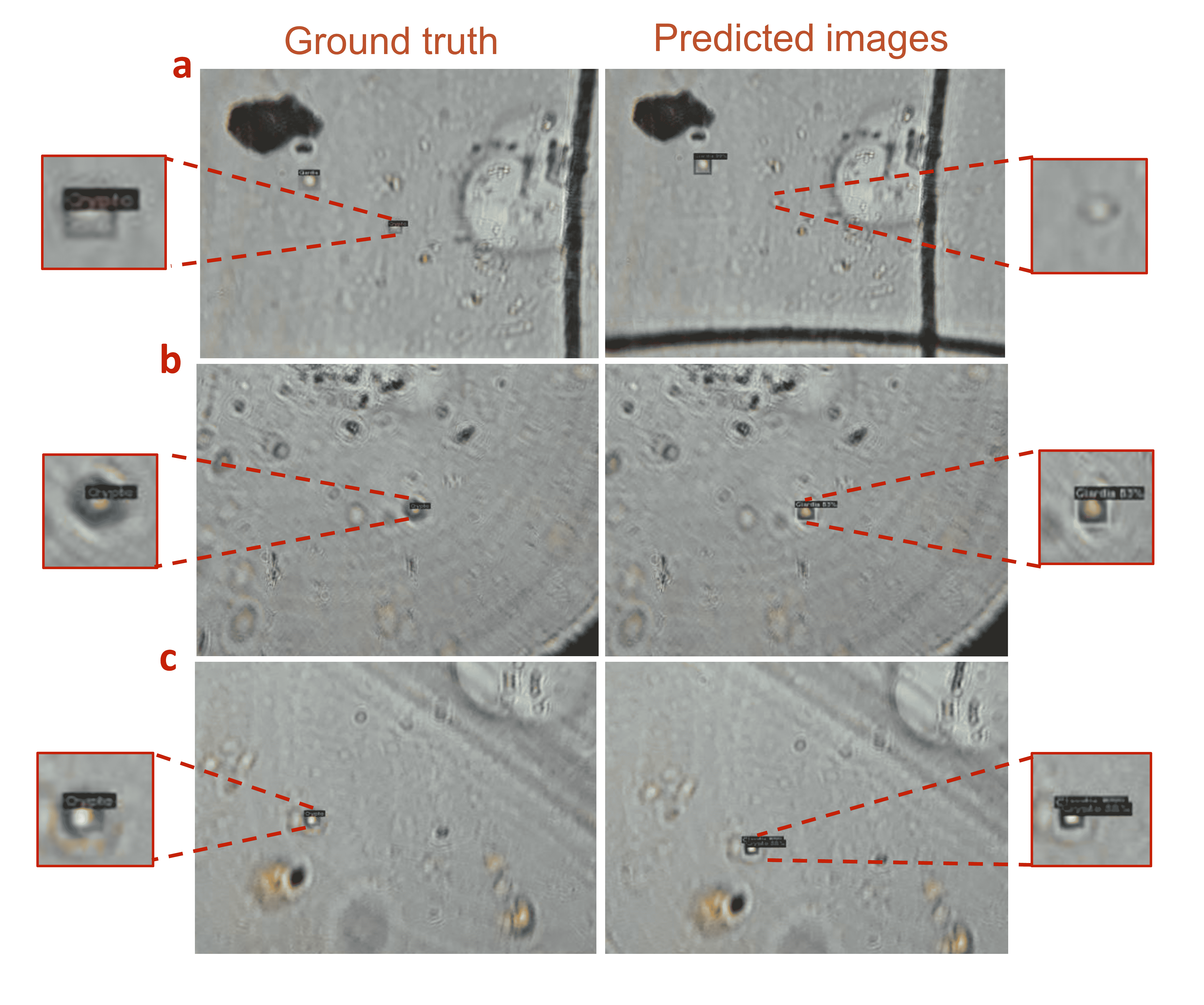}
    \caption{Three pairs of ground truth and predicted images (i.e., a, b, and c) showing error in prediction for vegetable sample test set taken from smartphone microscopy.  (\textbf{a}) Prediction missed due to blurry parasite, (\textbf{b}) False prediction due to similar size, and (\textbf{c}) Multiple predictions for a single object. For this illustration, Faster RCNN was used to detect parasites.}
    \label{fig:problem_smartphone}
\end{figure*}

The results show that for the same range of training samples, models perform better on reference samples than vegetable samples, brightfield microscopes than smartphone microscopes, \emph{Giardia} than \emph{Cryptosporidium} cysts.
Vegetable samples have debris similar to \emph{Giardia} and \emph{Cryptosporidium} (\autoref{fig:example_images}), making the task more difficult.
Smartphone images are more textured and noisy.
Additionally, due to the curvature effect of the ball lens in the smartphone microscope, the objects get stretched toward the peripheral regions and appear bigger than those at the center of the image.
In such cases, the models could falsely predict \emph{Giardia} as \emph{Cryptosporidium} and vice versa, as shown in \autoref{fig:problem_smartphone}. 
Some parasites on the smartphone images seem to be blurry, which causes false negatives as shown in \autoref{fig:problem_smartphone}a.
\autoref{fig:problem_smartphone}c illustrates a scenario of multiple predictions where both classes were predicted for a single object. 
However, it was only observed when using RetinaNet. 
In other models, the problem was eliminated by increasing the confidence threshold level for the prediction. 
\emph{Giardia} was better predicted by all four models used in the study, possibly due to its larger size and ellipse shape. 
At the time this article was published, to the best of our knowledge, no other work employed automation in sapphire ball lens-based smartphone microscopy to detect \emph{Giardia} and \emph{Cryptosporidium}. Some similar works are presented as follows: \cite{luo2021deep} reported a combined average sensitivity (or recall) of 0.974 for both the parasites using deep learning-enabled imaging flow cytometry; however, they stated the ML model did not work well in very different unseen data. \cite{koydemir2015rapid} presented a fluorescent-based smartphone microscopy system integrated with a machine learning algorithm to detect \emph{Giardia} that achieved a sensitivity of ~0.840 on water samples, reporting a prediction time of 2 minutes. \cite{ligda2020cryptosporidium} implemented Linear Discriminant Function Analysis (LDFA) to quantify the \emph{Giardia} and \emph{Cryptosporidium} with an accuracy of 0.690 and 0.750, respectively, on water samples using Olympus fluorescence microscope. 
Note that, among all these works, the smartphone-based microscopy system we have used is the cheapest, costing only \$15 - excluding the price of the smartphone \citep{shrestha2020smartphone}.

This work shows the feasibility and promise of integrating deep learning-based automated models into brightfield and smartphone microscopes, especially in resource-constrained areas where experts are not readily available.
Although the models performed better than non-experts in \emph{Cryptosporidium} cysts, the F1-scores for the models are still relatively low in test sets.
Future work requires collecting a much larger dataset, which will improve the scores.
Similarly, human experts can commit errors during annotation; therefore, it would be interesting to assess the inter- and intra-operator variability among the experts by annotating a certain subset and comparing this variability against AI models.
A larger dataset can be used for self-supervised pretraining on unlabeled samples, followed by supervised fine-tuning on the annotated datasets to get better object detection performance.
Moreover, in this study, we have not combined reference and vegetable sample images to train the models. 
The mixed dataset could be used in future works hoping for better performance and domain adaptation. Also, since the images obtained from smartphone microscopy were typically noisy, it would be beneficial to explore various image enhancement methods such as adaptive filters, Gaussian filters, deblurring techniques like Weiner deconvolution, Super-Resolution Convolutional Neural Network (SRCNN) \citep{dong2015image}, or deep-learning-based method \citep{zhao2020new} before training. Future works can also focus on developing robust models in detecting tiny \textit{Cryptosporidium} cysts. Additionally, defining the region of interest on the images to discard any stretched portions and applying domain adaptation techniques as suggested by \citep{becker2014domain} and \citep{ farahani2021brief} to transform smartphone microscopy images into images resembling brightfield microscopy images could further improve the performance of the models.

In our experiments, YOLOv8s performed decently in the real vegetable samples. 
Considering its lightweight architecture (22 MB) and good performance in detecting the cysts, YOLOv8s shows potential to be deployed on mobile devices without the need for a server \citep{kuznetsova2021YOLOv5, chen2021ship}.

This work provides the first step and shows the feasibility of a low-cost smartphone-based automated detection of (oo)cysts or other microorganisms in vegetables, water, stool, or other food products without needing an expert. 
More than specific deep learning models to choose from, future work should focus on larger datasets or semi-supervised approaches and designing experiments in prospective settings to compare against non-experts and experts for diagnostic end-points.


\acks{We thank Rabin Adhikari for helping us with formatting.
The images used in this research were taken with support from NAS and USAID (to BG and BBN) through Partnerships for Enhanced Engagement in Research (PEER) (AIDOAA-A-11-00012). 
The opinions, findings, conclusions, or recommendations expressed in this article are those of the authors alone and do not necessarily reflect the views of USAID or NAS. Suprim Nakarmi is currently affiliated with the University of South Dakota. 
}

%
\ethics{The work follows appropriate ethical standards in conducting research and writing the manuscript, following all applicable laws and regulations regarding the treatment of animals or human subjects.}

\coi{The authors declare no conflicts of interest.}


\bibliography{bibliography}


\clearpage
\appendix
\section{Supplementary Information}

\subsection{Implementation Details}
\label{sec:imp_details}
The networks were implemented in Python $3.10.6$.
The open-source object detection library, mmdetection \citep{chen2019mmdetection}, was used to execute Faster RCNN, RetinaNet, and Deformable DETR. Note that the backbone network of ResNeXt101 accompanied by Feature Pyramid Network (FPN), ResNet101, and ResNet50 were used for Faster RCNN, RetinaNet, and Deformable DETR, respectively.
Similarly, YOLOv8s was implemented by cloning the repository of Ultralytics \citep{yolov8s2023}. For Faster RCNN, RetinaNet, and Deformable DETR the shortest edge was resized with sharp edge lengths of 640, 472, 704, 736, 768, and 800, and a random horizontal flip with a probability of $0.5$ was used during data augmentation. For YOLOv8s, random horizontal flips were applied along with mosaic augmentation with probability $0.5$ and 1.0, respectively, for reference and vegetable sample images.
We adapted the learning rate, iterations, and number of classes for all three models during training using empirical experiments to achieve optimal performance.
Additionally, for RetinaNet, the \textit{Focal Loss's} alpha and gamma parameters were adjusted to improve the results from the default settings.
The hyperparameters used for these three models are summarized in \autoref{tab:models-paras}. 


\begin{table*}[htb!]
\centering
\caption{Eight detection models and hyperparameter details. $W/N$: Warm-up/Maximum Iteration; $LR$: Learning Rate; $FPN$: Feature Pyramid Network; $PPI$: Proposals per image; all other hyperparameters were left as default in mmdetection implementation.}
\label{tab:models-paras}
\begin{tabular}{|l|l|l|l|l|l|}
\hline
\textbf{Microscope} & \textbf{Detector} &\textbf{Backbone} &\textbf{W/N} & \textbf{LR} & \textbf{Other}\\
\hline
Brightfield & Faster RCNN & ResNeXt101 & 1200/1500 & $0.001$ & FPN, PPI = $64$ \\
             & RetinaNet &ResNet101  & 800/1200  & $0.001$ & $\alpha=0.93, \gamma=1$ \\
             & YOLOv8s & CSPDarknet & 3/100 & $0.01$ & batch size=16 \\
             & Deformable DETR & ResNet50 & - & $0.001$ & batch size =16, \\
             &&&&&epoch=100 \\
\hline
Smartphone  & Faster RCNN &ResNeXt101 & 1500/2000 & $0.01$ & FPN, PPI = 64\\
            & RetinaNet &ResNet101 & 1200/1500 & $0.001$ &  $\alpha=0.99, \gamma=1.7$ \\
            & YOLOv8s & CSPDarknet & 3/200 & $0.001$ & batch size = 16\\
            & Deformable DETR & ResNet50 & - & $0.01$ & batch size=16, \\
            &&&&&epoch=110 \\ 
\hline
\end{tabular}
\end{table*}

\subsection{Statistical Analysis}
\label{sec:stat}
The Quantile-Quantile (Q-Q) plot was used to check if the data were normally distributed.
Since the data were not normally distributed, we selected paired Wilcoxon signed-rank test \citep{woolson2007wilcoxon} to test the significance between the predictions.
A p-value of less than $0.05$ (i.e., $5$ $\%$) was considered significant.
We have assumed total images as the sample size (i.e., sample size = $193$) and the count of parasites on each image as the scores.
Since the sample size was large (i.e., $n > 50$), we used normal approximation on the Wilcoxon signed-rank test.
Here, normal approximation does not mean the data distribution is normal, but the Wilcoxon signed rank test statistic is assumed to be approximately normal.
We used scipy\footnote{\url{https://docs.scipy.org/doc/scipy/reference/generated/scipy.stats.wilcoxon.html}} to perform the Wilcoxon signed-rank test.

The null hypothesis is that there is no difference between the cysts-count predictions between the non-experts and models.
The p-values are provided in \autoref{tab:giardia_significance} for \emph{Giardia} and \autoref{tab:crypto_significance} for \emph{Cryptosporidium}.
We observed significantly different results among all the non-experts and between Non-expert $2$ and models. 
All three models significantly outperformed the non-experts for the detection of \emph{Cryptosporidium}.
However, for \emph{Giardia}, only Faster RCNN had a statistically significant difference in performance compared to Non-expert 1 and Non-expert 2.
In contrast, RetinaNet had a statistically significant difference in performance compared to Non-expert 3. 
Faster RCNN and RetinaNet were significantly better than non-expert human $3$ in predicting \emph{Giardia} and \emph{Cryptosporidium}.

\begin{table*}[h!]
\centering
\caption{Table showing the p-values for \emph{Giardia} - using Wilcoxon signed rank test - among expert humans, non-expert humans, and AI. (p-value threshold of 0.05).
}
\label{tab:giardia_significance}
\resizebox{\textwidth}{!}{%
\begin{tabular}{c|c|c|c|c|c|c|c|}
\cline{2-8}
                                   & Expert   & Non-expert 1 & Non-expert 2 & Non-expert 3 & Faster RCNN & RetinaNet & YOLOv8s \\ \hline
\multicolumn{1}{|c|}{Non-expert 1} & \textbf{1.64E-04} &              &              &              &             &     &      \\ 
\multicolumn{1}{|c|}{Non-expert 2} & \textbf{4.72E-04} & \textbf{1.29E-08}     &              &              &             &       &    \\ 
\multicolumn{1}{|c|}{Non-expert 3} & \textbf{1.98E-04} & \textbf{4.04E-08}     & 7.62E-01     &              &             &    &       \\ 
\multicolumn{1}{|c|}{Faster RCNN}  & \textbf{2.19E-09} & 2.60E-02     & \textbf{5.71E-13}     & \textbf{7.88E-12}     &             &       &    \\ 
\multicolumn{1}{|c|}{RetinaNet}    & \textbf{1.08E-09} & 2.59E-02     & \textbf{1.02E-13}     & \textbf{2.60E-12}     & 7.79E-01    &     &      \\ 
\multicolumn{1}{|c|}{YOLOv8s}      & \textbf{1.10E-13} & \textbf{1.57E-05}     & \textbf{5.18E-17}     & \textbf{6.42E-15}     & \textbf{4.50E-05}    & \textbf{7.74E-05}&\\ 
\multicolumn{1}{|c|}{Deformable DETR}      & \textbf{4.09E-10} & {1.16E-3}     & \textbf{1.16E-14}    & \textbf{6.04E-21}     & \textbf{3.56E-10}    & \textbf{1.40E-24} & {6.69E-3}  \\ \hline
\end{tabular}%
}
\end{table*}

\begin{table*}[h!]
\centering
\caption{Table showing the p-values for \emph{Cryptosporidium} - using Wilcoxon signed rank test - among expert humans, non-expert humans, and AI. (p-value threshold of 0.05).
}
\label{tab:crypto_significance}
\resizebox{\textwidth}{!}{%
\begin{tabular}{c|c|c|c|c|c|c|c|}
\cline{2-8}
                                   & Expert   & Non-expert 1 & Non-expert 2 & Non-expert 3 & Faster RCNN & RetinaNet & YOLOv8s \\ \hline
\multicolumn{1}{|c|}{Non-expert 1} & \textbf{1.14E-04} &              &              &              &             &        &   \\
\multicolumn{1}{|c|}{Non-expert 2} & \textbf{3.03E-21} & \textbf{5.52E-25}     &              &              &             &    &       \\
\multicolumn{1}{|c|}{Non-expert 3} & \textbf{9.37E-03} & \textbf{8.51E-10 }    & \textbf{8.88E-21}     &              &             &    &       \\
\multicolumn{1}{|c|}{Faster RCNN}  & \textbf{5.60E-04} & 2.62E-02     & \textbf{1.86E-24}     & \textbf{4.33E-09}     &             &     &      \\
\multicolumn{1}{|c|}{RetinaNet}    & \textbf{1.37E-04} & 1.56E-01     & \textbf{3.52E-26}     & \textbf{6.40E-10}     & 3.23E-01    &      &     \\
\multicolumn{1}{|c|}{YOLOv8s}      & 6.67E-02 & \textbf{8.97E-04}     & \textbf{2.01E-23}     & \textbf{1.82E-06}     & \textbf{5.43E-03}    & \textbf{6.33E-04} &  \\ 
\multicolumn{1}{|c|}{Deformable DETR}      & {7.58E-2} & \textbf{3.34E-08}     & \textbf{2.16E-22}     & \textbf{5.74E-08}     & \textbf{8.00E-08} & {7.04E-2} & \textbf{3.63E-10}  \\ 
\hline
\end{tabular}%
}
\end{table*}

\subsection{Choice of Thresholds}
\label{sec:thresholds_choice}
\begin{figure*}[htb]
  \centering
  \includegraphics[width=\textwidth]{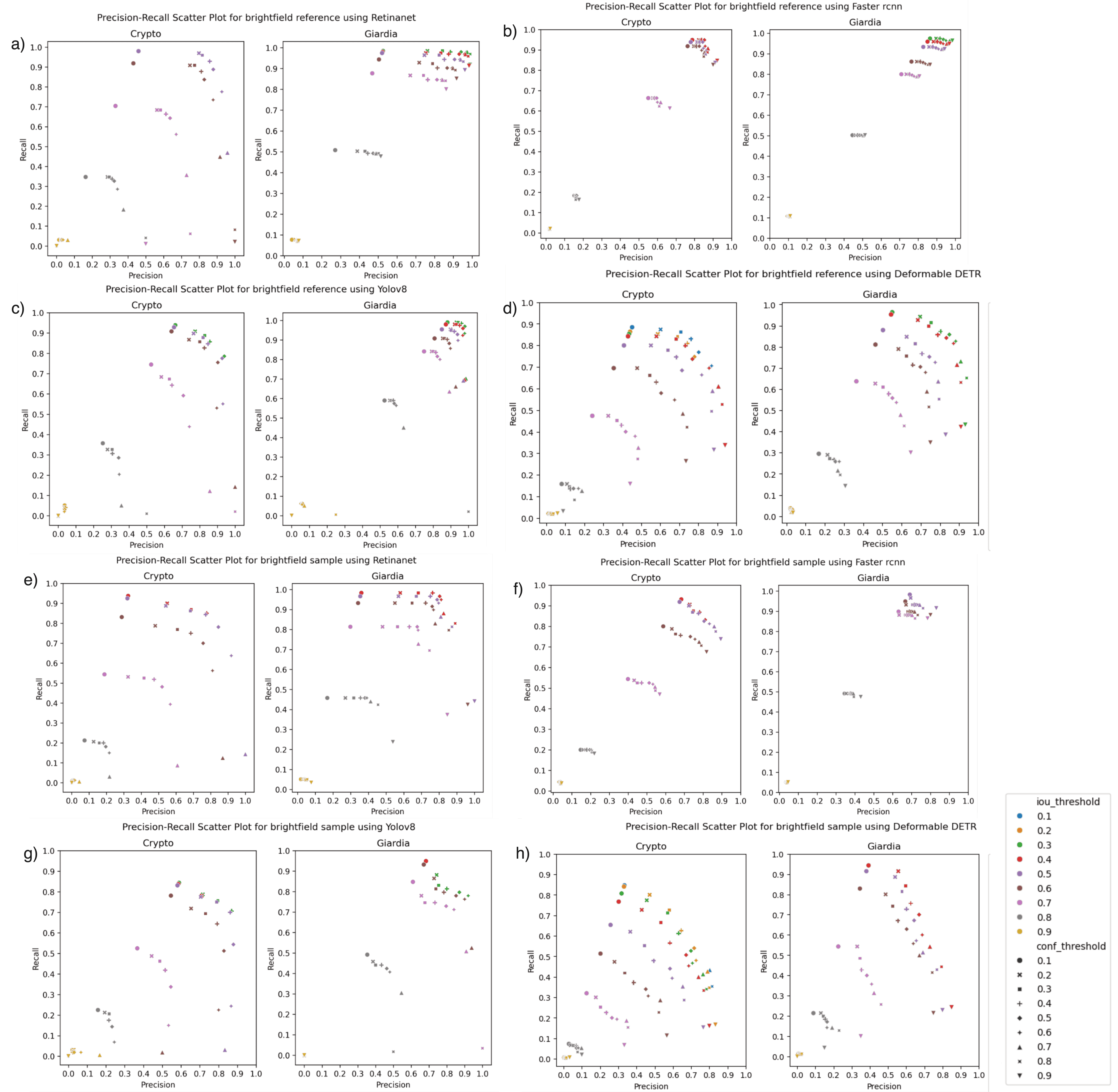}
  \caption{Precision-Recall scatterplot for different values of iou threshold and confidence score for brightfield reference samples and vegetable samples on four detection models. Each subfigure displays scatterplots for (a) brightfield reference sample using RetinaNet, (b) brightfield reference sample using Faster RCNN, (c) brightfield reference sample using YOLOv8s, (d) brightfield reference sample using Deformable DETR, (e) brightfield vegetable sample using RetinaNet, (f) brightfield vegetable sample using Faster RCNN, (g) brightfield vegetable sample using YOLOv8s, and (h) brightfield vegetable sample using Deformable DETR.}
  \label{fig:brightfield_thres}
\end{figure*}

\begin{figure*}[htb]
  \centering
  \includegraphics[width=\textwidth]{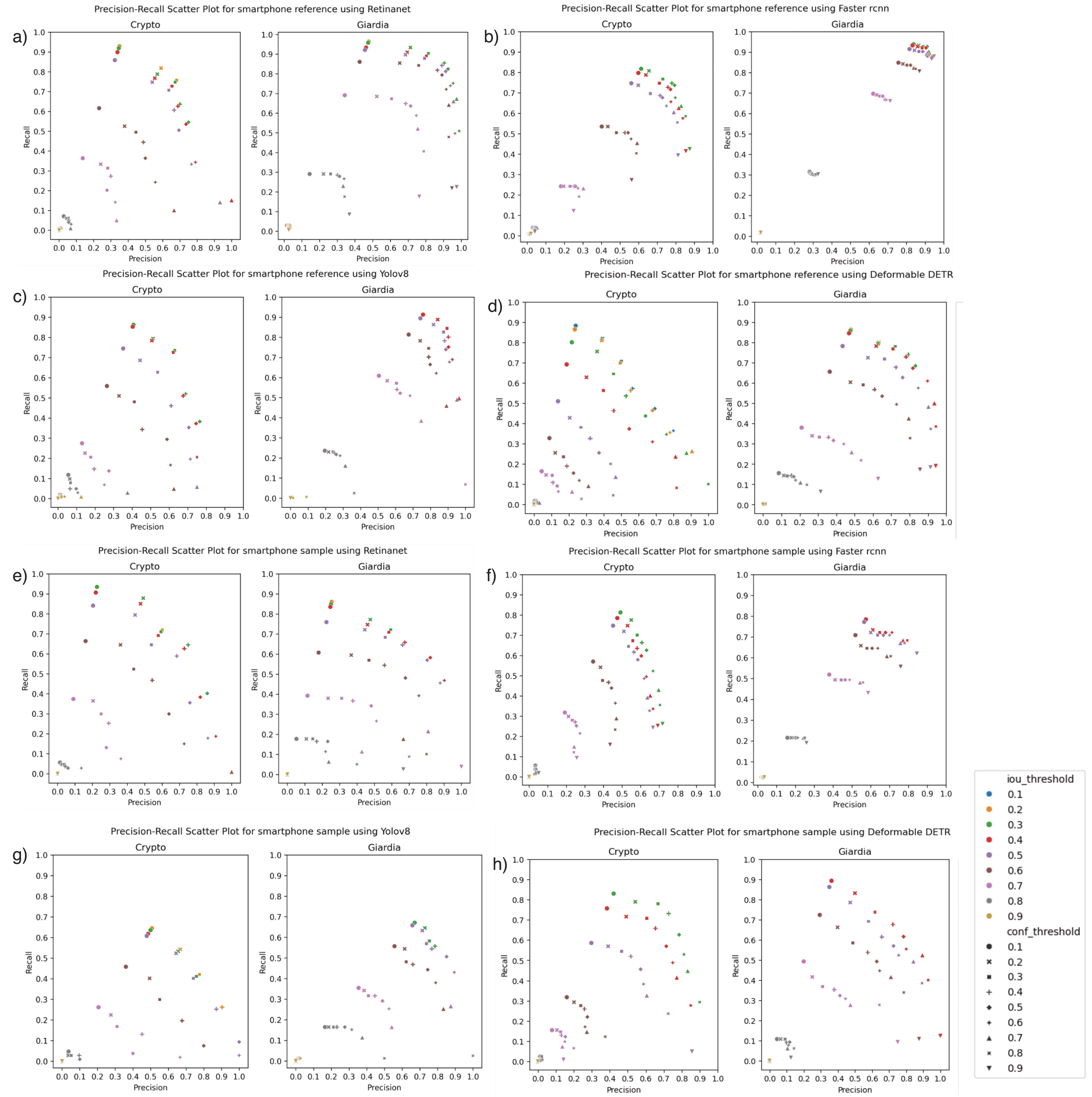}
  \caption{Precision-Recall scatterplot for different values of iou threshold and confidence score for smartphone reference samples and vegetable samples on four detection models. Each subfigure displays scatterplots for (a) smartphone reference sample using RetinaNet, (b) smartphone reference sample using Faster RCNN, (c) smartphone reference sample using YOLOv8s, (d) smartphone reference sample using Deformable DETR, (e) smartphone vegetable sample using RetinaNet, (f) smartphone vegetable sample using Faster RCNN, (g) smartphone vegetable sample using YOLOv8s, and (h) smartphone vegetable sample using Deformable DETR.}
  \label{fig:smartphone_thres}
\end{figure*}

To calculate the precision, recall, and F1-score of the object detection models, we plotted the precision-recall of the respective detectors for different values of iou thresholds and confidence scores, ranging from 0.1 to 1 with a step size of 0.1, as shown in \autoref{fig:smartphone_thres} and \autoref{fig:brightfield_thres}.
Then, we selected the thresholds by observing the respective plots to get the best precision and recall.

\end{document}